# Collision Cross Sections of the J = 2 ← 1 Rotational Transition of $CF_3CCH$ due to Higher Order Interactions

H Maeder[1], J Galica[2], E Mis-Kuzminska[2] and S. Gierszal[2]

[1]Institut fur Physikalische Chemie  der Universitat Kiel, Olshausenstrasse 40,
D-24098 Kiel,Germany
[2]Institute of Molecular Physics, Polish Academy of Sciences,
ul. M. Smoluchowskiego 17, 60-179 Poznań, Poland

E-mail: maeder@phc.uni-kiel.de, galica@ifmpan.poznan.pl

**Abstract**
The collision cross section of the rotational transition J = 2 ← 1 of 3,3,3
-trifluoropropyne, $CF_3CCH$, caused by rare gas perturbers has been determined by
investigating transient emission signals of molecular gas samples. From analysis
of the pressure dependence of the width of the rotational line J = 2←1 pressure
broadening parameters have been derived for the pure gas and for mixtures with
the rare gases He, Ne, Ar, Kr and Xe. The pressure shift parameter for the pure
gas $\delta_s/p$ = 29.03(12) kHz/Pa also has been obtained. Calculations based on the
Anderson-Tsao-Curnutte theory using induction and dispersion interactions for
the description of collisions of $CF_3CCH$ with He, Ne, Ar, Kr and Xe, respectively,
are in qualitative agreement with the experimental results.

## 1. Introduction

Trifluoromethylacetylene (3,3,3-trifluoropropyne), $CF_3CCH$, is an attractive symmetric top molecule, the rotational spectrum of which has been measured for the first time in 1951 [1]. The microwave rotational spectra of the ground state of $CF_3CCH$ were measured by several groups [1 - 6]. Several papers are devoted to the lowest degenerate exited state $v_{10}$=1 [1, 4,5, 7,8,9] as well as to the higher excited states $v_{10}$ = 2 [4,7,10, 11], $v_{10}$ = 3 [4,7,12,13], and $v_{10}$ = 4 [13]. The structure of the $CF_3CCH$ molecule was determined by combining  the results from microwave absorption spectroscopy and electron diffraction technique [7], as well as with investigations under high resolution using a pulsed-nozzle Fourier transform microwave spectrometer [3]. Investigations on this molecule by microwave spectroscopy were also carried out studying rotational energy transfer by four level, modulated double resonance [14], triple-resonance phenomena by two-dimensional microwave Fourier transform spectroscopy [15], and measuring the Stark effect by Fourier transform microwave  spectroscopy (FTMW) [6]. The latter technique has been successfully used to determine $T_2$- type rotational relaxation times for 3,3,3-trifluoropropyne in the gaseous phase [16]. The rotational transitions:  (J'←J) = (1←0), (2←1), (3←2) of the $CF_3CCH$ molecule have been investigated and linewidth-parameters ($\Delta\nu/p$) for self-, $H_2$ and He-collisions at room temperature were determined.
In this paper we present the experimental results obtained by Fourier transform microwave spectroscopy in which linewidth measurements were performed to determine the collision cross sections of 3,3,3-trifluoropropyne interacting with He, Ne, Ar, Kr and Xe atoms. The purpose of these investigations is to provide quantitative information on the molecular cross sections and on the intermolecular interactions which characterize the collision partners in the selected mixtures of gases .
When a dipolar absorption line is  broadened due to collisions of the absorber  molecule with a noble gas atom, the intermolecular forces responsible for the broadening would be the induction forces (dipole - induced dipole, quadrupole - induced dipole, dipole - induced quadrupole), the dispersion and exchange forces.

The theory of pressure broadening in the microwave region was given by Anderson, Tsao and Curnutte (ATC) [17,18], and extended for the dispersion and exchange interactions by Krishnaji [19 - 22]. We use these theories for the interpretation of our experimental results in order to determine from the linewidth parameters the collision cross sections of $CF_3CCH$ interacting with noble gas atoms.

## 2. Experimental

The measurements of microwave absorption were carried out with a Fourier transform microwave spectrometer (FTMW) as described elsewhere [23]. The experiment was carried out for the $J = 2 \leftarrow 1$ rotational transition in the ground vibrational state of $CF_3CCH$. The pressure dependent transition frequency is around 11511.8 MHz. The sample of $CF_3CCH$ was used after vacuum distillation in the experiment. Foreign gases of research purity were used for the investigation of the mixtures with He, Ne, Ar, and Kr. The pressure ranges of the gases measured in the absorption cell were dependent on the intensity of the transition and in the range between: 0.1 to 2.4 Pa for self-broadening and 0 to 6 Pa for foreign gases at the mixtures with 1 Pa fixed partial pressure of $CF_3CCH$. The sample pressure of trifluoropropyne as well as the total pressure of the gas mixtures was measured with an absolute pressure gauge MKS Baratron 170M calibrated to better than 1% accuracy. During the recording of the absorption line the gas pressure is maintained to within 0.03 Pa. All measurement were performed at room temperature ($298 \pm 1$ K).

## 3. The Model

Interactions which occur during molecular collisions can be described as electrostatic, inductive, dispersive and exchange. Collisions of polar molecules with noble gas atoms—molecules are a special case, involving only inductive, dispersive and exchange interactions. The inductive processes may be detailed as (i) dipole–induced dipole ($\mu_1\alpha_2\mu_2$); (ii) dipole–induced quadrupole ($\mu_1\alpha_2q_1$); (iii) quadrupole–induced dipole ($q_1\alpha_2\mu_1$) where $\alpha_2$ is the mean polarizability of the perturber molecule. The dipole-induced dipole, similarly as the dispersive interactions, vary as $b^{-10}$, where b is the collision parameter; the rest of the inductive processes which involve the quadrupole moment $q_1$ of the absorber molecule are proportional to $b^{-12}$. The theoretical description of inductive, dispersive and exchange interactions, based on the Anderson and Tsao–Curnutte theory [17,18], was developed by Krishnaji and Shrivastava [19 - 22].

## 4. Theory

The collision cross section $\sigma$ is defined with the integral:

$$\sigma = \int_0^\infty 2\pi b S(b) db, \quad (1)$$

where $b$ is the impact parameter and $S(b)$ is the weight factor indicating the differential probability of collision induced transitions which depends upon the interaction Hamiltonian of both the absorber and the perturber. In the experiment the dipolar absorption rotational line $J = 2 \leftarrow 1$ of $CF_3CCH$ is broadened due to collisions with noble gas atoms, hence the interaction forces responsible for the broadening are induction and dispersion forces.

The expression of $S(b)$ for the dispersion (DSP) and induction (IND) interactions can be written as the sum

$$S(b) = S_{DSP} + S_{IND} \quad (2)$$

where:

$$S(b)_{DSP} = \frac{A_{DSP}}{b^{10}} F(k), \quad (3)$$

$$S(b)_{IND} = \frac{A_{IND}}{b^{10}} F(k) \quad (4)$$

$$A_{DSP} = \frac{21\pi^3}{10240\hbar^2 kNT} \left[ \frac{I_1 I_2}{I_1 + I_2} (\alpha_1^\| - \alpha_1^\perp)\alpha_2 \right]^2 M, \quad (5)$$

$$A_{IND} = \frac{21\pi^3}{540\hbar^2 kNT}(\alpha_2 \mu_1^2)^2 M, \tag{6}$$

$$F(k) = \sum_{J'_i} Q(J_i J'_i) g(k) + \sum_{J'_f} Q(J_f J'_f) g(k) + B, \tag{7}$$

$$B = (-1)^{J_i+J_f} 2[(2J_i+1)(2J_f+1)Q(J_iJ_i)Q(J_fJ_f)]^{1/2} W(J_iJ_fJ_iJ_f,12) \tag{8}$$

with $\mu_1$: dipole moment of absorber, $I_1$, $I_2$: ionisation energies of the colliding molecules/atoms, $J_i$ $J_f$: initial and final states of absorber, $\alpha_1^{\parallel}$, $\alpha_1^{\perp}$: components of the polarizability along and perpendicular to the symmetry axis of the absorber, $\alpha_2$: mean polarizability of the perturber, $T$: absolute temperature, $M$: reduced mass of colliding molecules, $Q(J_iJ_i')$, $Q(J_fJ_f')$: quadrupole transition probabilities, $W(J_iJ_fJ_iJ_f,12)$: Racah coefficient, $N$: Avogadro's constant, $g(k) = g([b/v]|\omega|)$: function tabulated by Tsao - Curnutte [18], $v$: relative velocity, $\omega$: change in the rotational energy of absorber.

## 5. Results and discussion

From the observed time-dependence of the transient emission signal for a fixed pressure of the gas, the collisional dephasing time ($T_2$) is obtained [24]. $T_2$ is often referred to as "coherence decay time" since it describes a loss of coherent polarization of the sample due to intermolecular interaction. The half-width $\Delta \nu$ at half-height of the Lorentzian line at low radiation power is related to $T_2$ by

$$\Delta \nu = \frac{1}{2\pi T_2} \tag{9}$$

The dependence of the linewidth of the rotational transition $J = 2 \leftarrow 1$ of trifluoropropyne on the pressure $p_1$ of the perturber gas is determined from the formula,

$$\Delta \nu = C_w p_0 + C_w(x) p_1 \tag{10}$$

where $C_w$ is the pressure self-broadening coefficient, $C_w(x)$ is the pressure broadening coefficient of the gas mixture, $p_0$, $p_1$ are the partial pressures of $CF_3CCH$ and perturber gas, respectively. The value of $C_w p_0$ is determined from the measured linewidth for the pure $CF_3CCH$ at the initial pressure $p_0$. Gradually, the perturbing gas is added and the linewidth $\Delta \nu$ measured, hence, the difference $\Delta \nu - C_w p_0$, resulting from collisions with the perturber, can be calculated and the foreign gas pressure coefficient.
From the plot of $\Delta \nu$ vs. the pressure of the pure gas 'figure 1', the pressure self-broadening coefficient $C_w$ for the rotational transition $J = 2 \leftarrow 1$ of trifluoropropyne was determined from a linear regression analysis to $C_w = 343.8(22)$ kHz/Pa.

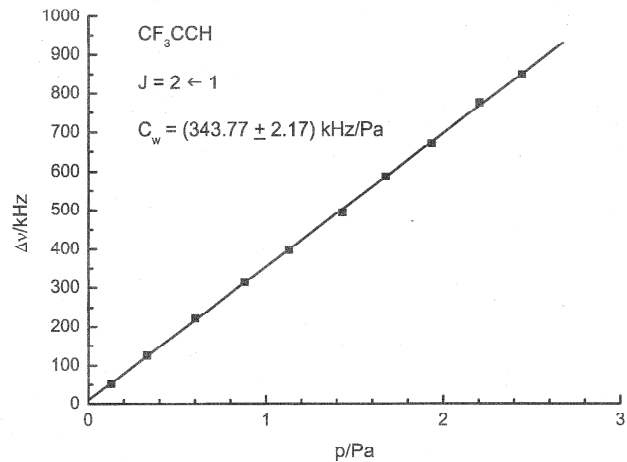

**Figure 1.** The linewidth (Δν) of the J = 2 ← 1 rotational transition of CF$_3$CCH as a function of the pure gas pressure (p).

This value is in agreement with that reported earlier, $C_w$ = 339.0(75) kHz/Pa (45.2(10) MHz/Torr) [16]. The centre frequency of of the transition ν$_0$ = 11511.7909(30) MHz has been obtained by extrapolation of experimental results to the zero pressure 'figure 2'. It corresponds with number 11511.7962(24) MHz referred in the literature [16].

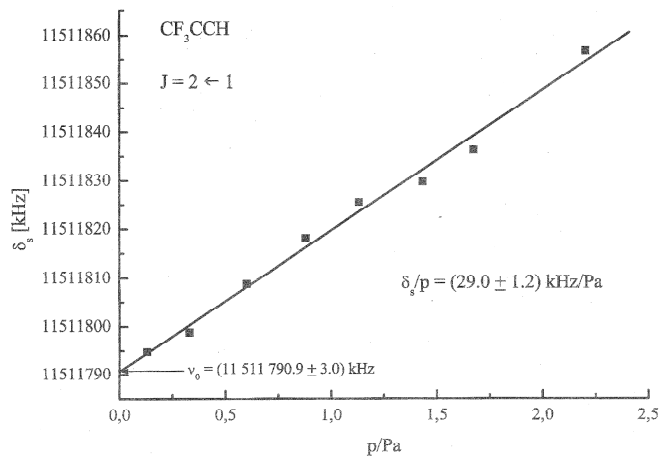

**Figure 2.** The center frequencies (ν) of the J = 2 ← 1 rotational transition of CF$_3$CCH versus the pure gas pressure (p).

The experimentally measured centre frequencies of the J = 2 ← 1 rotational transition versus pressure p is displayed in 'figure 2'. The value of the measured pressure shift parameter is $C_s$ = 29.0(12) kHz/Pa.
For CF$_3$CCH mixtures with He, Ne, Ar, Kr and Xe, respectively, the differences Δν – $C_w p_0$ are plotted against the partial pressure of perturbing gas p$_1$ = p – p$_0$, see for example mixtures with He, Ne and Xe in 'figure 3'. The pressure broadening coefficient $C_w(x)$ can subsequently by obtained as the least squares linear fit and amounts to : 32.4(14), 24.5(10), 28.6(14), 25.5(12) and 25.9(12) kHz/Pa for mixtures of CF$_3$CCH with He, Ne, Ar, Kr and Xe, respectively.

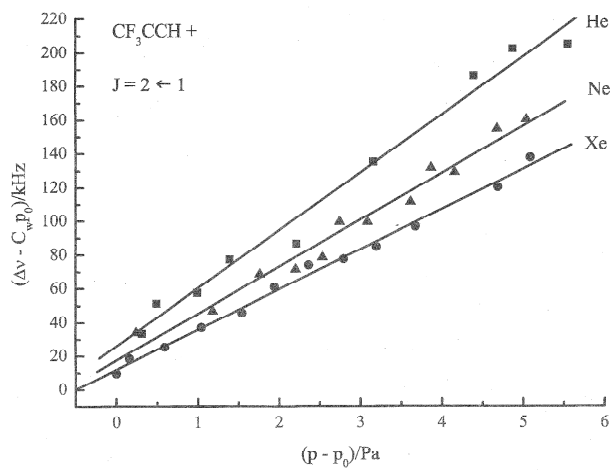

**Figure 3.** Differences Δν – $C_w p_0$ for the linewidth of the J = 2 ← 1 rotational transition of CF$_3$CCH as a function of the partial pressure (p – p$_0$) of perturbing gases He, Ne and Xe, respectively.

The collision cross section σ(x) for the molecules perturbed by foreign gas molecules is given as follows [25]:

$$\sigma(x) = \frac{2\pi kT}{v} C_w(x) \tag{11}$$

where $C_w(x)$ is the experimentally obtained foreign gas broadening parameter, x = He, Ne, Ar, Kr, Xe; k the Boltzman constant, T the temperature and v the mean relative velocity of the collision partners. Alternatively, the pressure broadening coefficients $C_{wcalc}(x)$ and collision cross-sections $\sigma_{calc}$ were determined theoretically for the investigated mixtures of trifluoropropyne with rare gases. The quantities where determined on the basis of the Anderson- and Tsao-Curnutte- (ATC) theory within the limits of dispersive and inductive interaction only. The values of the parameters which were used in the calculations are listed in Table 1.

Table 1. Polarizability and ionisation energy of $CF_3CCH$ and perturber atoms.

|  | $CF_3CCH$ | He | Ne | Ar | Kr | Xe |
|---|---|---|---|---|---|---|
| average polarizability [cm$^3$] $\alpha_1, \alpha_2 *10^{24}$ | 2.595[b] | 0.205[a] | 0.396[a] | 1.641[a] | 2.484[a] | 4.044[a] |
| difference of polarizability $\alpha_1^{\parallel} - \alpha_1^{\perp} *10^{24}$ [cm$^3$] | 5.818[b] |  |  |  |  |  |
| ionisation energy $I_1, I_2$ [eV] | 12.68[b] | 24.59[a] | 21.56[a] | 15.76[a] | 14.00[a] | 12.13[a] |
| Dipole moment $\mu$ [D] | 2.317[c] |  |  |  |  |  |

[a] – from "Handbook of Chemistry and Physics", Lide, 78$^{th}$ edition (1997-1998)
[b] – program MINDO/3 POLAR PRECISE
[c] – ref. [6]

The theoretical results for $C_{wcalc}$ and $\sigma_{calc}$ with consideration of the dispersive, inductive and dispersive-inductive interactions are contained in Table 2 together with the experimental values. In addition, the values of $b_0$, i.e. the numbers which satisfy the condition $S(b_0) = 1$, are included. In the calculations we assume that the function F(k) is independent of the energy levels structure of the collision partner and its value is denoted $F_0$ [25]. With these assumptions, and with the condition $b_0 \geq d$ (d: kinetic collision radius [26]), Odashima [25] and Kajita [27] have calculated the magnitude of the collision cross-section for inductive $\sigma_{ind}$ and dispersive $\sigma_{dis}$ interactions as follows:

$$\sigma_{ind} = \frac{5}{4}\pi B_{ind}^{\frac{1}{5}} \left(M\mu_1^4 \alpha_2^2\right)^{\frac{1}{5}} \tag{12}$$

$$\sigma_{dsp} = \frac{5}{4}\pi B_{dsp}^{\frac{1}{5}} \left[M\left(\frac{I_1 I_2}{I_1 + I_2}\right)^2\right]^{\frac{1}{5}} \tag{13}$$

From these relations (12, 13), it follows that the cross sections are proportional to $(M\alpha_2^2)^{1/5}$ for induction and to $\left[M\left(\frac{I_2}{I_1+I_2}\alpha_2\right)^2\right]^{\frac{1}{5}}$ for dispersion interactions. 'Figure 4' shows the measured cross section for collisions between $CF_3CCH$ molecules and rare gas atoms as a function of the quantity $(\alpha^2 M)^{\frac{1}{5}}$.

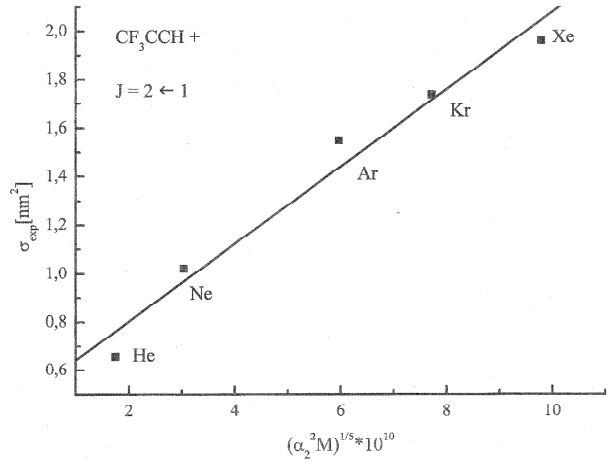

**Figure 4.** Experimentally obtained cross section of $CF_3CCH$ as a function of the quantity $(\alpha_2^2 M)^{1/5}$, where $\alpha_2$ is the polarizability of the perturber atom.

Similarly, 'figure 5' shows the measured collision cross section as a function of the quantity $\left[ M \left( \frac{I_2}{I_1 + I_2} \alpha_2 \right)^2 \right]^{\frac{1}{5}}$.

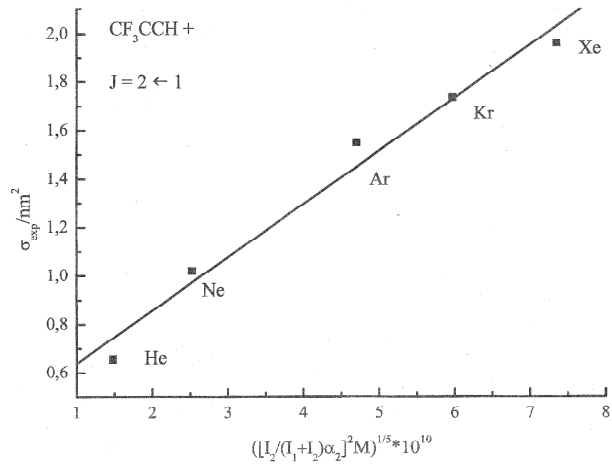

**Figure 5.** Experimentally obtained collision cross sections of $CF_3CCH$ as a function of the quantity $[(\alpha_2 I_2/(I_1+I_2))^2 M]^{1/5}$, where $I_1$, $I_2$ are the ionization energy of absorber and perturber, respectively.

Assuming that the two mechanisms of interaction occur at the same time, the cross section is [26]:

$$\sigma_{ind+dsp} = \frac{5}{4} \pi B_{dsp}^{\frac{1}{5}} \left[ 16 M \left( \mu_1^2 \alpha_2 \right)^2 + M \left( \frac{I_1 I_2}{I_1 + I_2} \gamma \alpha_1 \alpha_2 \right)^2 \right]^{\frac{1}{5}} \tag{14}$$

where: $\gamma = \dfrac{\alpha_1^{II} + \alpha_1^{\perp}}{\alpha_1}$ .

Consequently, the result of the inductive and dispersive interactions will be a cross-section proportional to the expression:

$$X = M^{\frac{1}{5}}\left[16\left(\mu_1^4\alpha_2^2\right) + \left(\frac{I_1 I_2}{I_1 + I_2}\gamma\alpha_1\alpha_2\right)^2\right]^{\frac{1}{5}} \qquad (15)$$

Experimentally obtained as well as theoretically calculated collision cross section of $CF_3CCH$ in collision with a rare gas atom plotted as a function of the quantity X is shown in 'figure 6'.

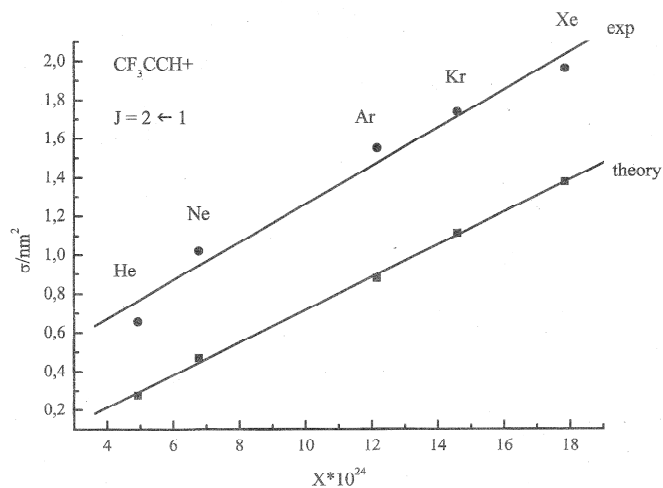

**Figure 6.** Experimentally and theoretically obtained collision cross section of $CF_3CCH$ as a function of X for various perturbers.

There is a linear relation between the cross section and the quantity X which represents the inductive and dispersive interaction. The theoretical values of $\sigma$ are calculated under the assumption that the interactions are inductive, dispersive and inductive-dispersive type. Generally the theoretical values are much smaller than the numbers obtained from the experiment (Table 2). It appears from the theoretical calculations, that dispersive interactions are the main cause of line broadening in collisions between the trifluoropropyne molecule and rare gas atoms.

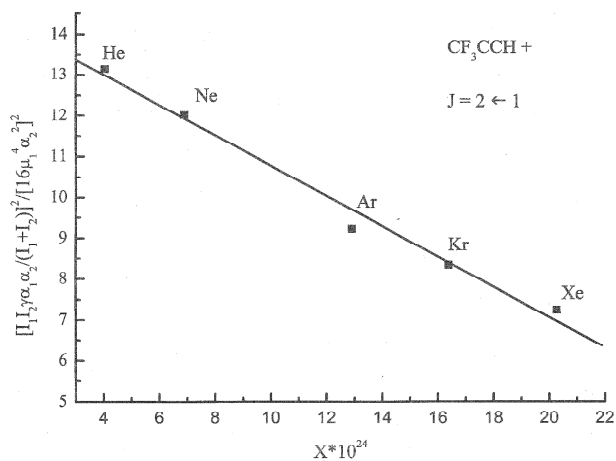

**Figure 7.** Dependence $[I_1 I_2 \gamma\alpha_1\alpha_2/(I_1+I_2)]^2/[16\mu_1^4\alpha_2^2]$ on the parameter $X = M^{1/5}\{[16\mu_1^4\alpha_2^2] + [I_1 I_2 \gamma\alpha_1\alpha_2/(I_1+I_2)]^2\}^{1/5}$ for various perturbers.

Table 2. Pressure broadening coefficients, $C_w$ [kHz/Pa], collision cross – sections $\sigma$ [nm$^2$], parameter $b_0$ [nm] determined for the J = 1 → 2 transition of $CF_3CCH$ in mixtures with noble gases at room temperature; $\sigma_{calc}$ – calculated collision cross-section, $\sigma_{exp}$ – calculated from Eq. 11

|  | Interactions | $C_w$ [kHz/Pa]<br>$\sigma$ [nm$^2$]<br>$b_0$ [nm] | He | Ne | Ar | Kr | Xe |
|---|---|---|---|---|---|---|---|
| Calc. | IND | $C_{wcalc}$<br>$\sigma_{calc}$<br>$b_0$ | 8.03<br>0.20<br>0.22 | 6.75<br>0.35<br>0.29 | 10.11<br>0.68<br>0.42 | 10.35<br>0.88<br>0.47 | 11.85<br>1.12<br>0.53 |
|  | DSP | $C_{wcalc}$<br>$\sigma_{calc}$<br>$b_0$ | 13.43<br>0.33<br>0.29 | 11.03<br>0.57<br>0.38 | 15.75<br>1.07<br>0.52 | 15.83<br>1.35<br>0.59 | 17.55<br>1.67<br>0.65 |
|  | IND+DSP | $C_{wcalc}$<br>$\sigma_{calc}$<br>$b_0$ | 13.64<br>0.28<br>0.29 | 11.22<br>0.47<br>0.38 | 16.08<br>0.88<br>0.53 | 16.21<br>1.11<br>0.59 | 18.01<br>1.37<br>0.66 |
| Exp. |  | $C_w$<br>$\sigma_{exp}$ | 32.4(14)<br>0.66(3) | 24.53(97)<br>1.02(4) | 28.65(142)<br>1.55(8) | 25.58(16)<br>1.74(1) | 25.95(120)<br>1.96(9) |

The relative magnitude of the two mechanism: dispersive interaction, proportional to $[\alpha_1\alpha_2\gamma I_1 I_2/(I_1+I_2)]^2$, is larger than the inductive which is characterised by the expression $16(\mu_1^2\alpha_2)^2$ for the range of perturbers what is demonstrated by the plot in 'figure 7'.

## 6. Conclusion

When a $CF_3CCH$ molecule collides with rare gas atom, the absolute values of the experimentally obtained cross section are significantly larger than those of calculated cross section in ATC theory. The experimentally obtained cross section depends linearly on the magnitude of X which defines the dispersion and dipole-induced dipole interactions. This means that the cross section is determined by the mass, polarizability and ionisation energy of the collision partner.

**Acknowledgement**

The authors thank Dr. U. Andresen and S. Jacobsen for help and discussions. J.G. is specially grateful to the Alexander von Humboldt Foundation for financial support in the form of a fellowship.

**References:**

[1] Anderson W E, Trambarulo R, Sheridan J and Gordy W 1951 *Phys. Rev.* **82** 58
[2] Martinache L, Ozier I and Bauder A C 1990 *J. Chem. Phys.* **92** 7128
[3] Cox P A, Ellis M C, Legon A C and Wallwork A 1993 *J. Chem.Soc.Farad.Trans.* **89** 2937
[4] Mills I M 1969 *Mol. Phys.* **16** 345
[5] Harder H , Gerke C, Maeder H, Cosleou J, Bocquet R, Demaison J, Papousek D and Sarka K 1994 *J. Mol. Spectrosc.* **167** 24
[6] Kasten W and Dreizler H 1984 *Z. Naturforsch.* **39a** 1003
[7] Shoolery J ., Shulman R. G, Sheehan W F, Schomaker V and Yost Don M 1951 *J. Chem. Phys.* **19** 1364

[8] Jaseja T S 1959 *Proc. Ind. Acad. Sci.A* **50** 108
[9] Grenier-Besson M L and Amat G 1962 *J. Mol. Spectrosc.* **8** 22
[10] Wotzel U, Maeder H, Harder H, Pracna P and Sarka K 2005 *Chem.Phys.* **312** 159
[11] Carpenter J H, Motamedi M and Smith JG 1995 *J. Mol. Spectrosc.* **171** 468
[12] Wotzel U, Maeder H, Harder H., Pracna P and Sarka K 2006 *J.Mol.Struct.* **780-781** 206
[13] Carpenter J H, Motamedi M and Smith J G 1995 *J. Mol. Spectrosc.* **174** 106
[14] Cohen J B and Wilson BE 1973 *J. Chem. Phys.* **58** 456
[15] Vogelsanger B 1990 *Z.Naturforsch.* **45a** 707
[16] Mehrotra S C, Dreizler H and Mäder H 1985 *J.Quant.Spectr.Radiat.Transfer* **34** 229
[17] Anderson P W 1949 *Phys. Rev.* **76** 647
[18] Tsao C J and Curnutte B 1962 *J.Quant.Spectr.Radiat.Transfer* **2** 42
[19] Krishnaji, Chandra S and Srivastava S L 1964 *J.Chem.Phys.* **41** 409
[20] Krishnaji and Srivastava S L 1964 *J.Chem.Phys.* **41** 2266
[21] Krishnaji and Srivastava S L 1965 *J.Chem.Phys.* **2** 1546
[22] Krishnaji and Srivastava S L 1965 *J.Chem.Phys.* **43** 1345
[23] Sarka K, Papousek D, Demaison J, Maeder H, and Harder H 1997
    *Advanced Series in Physical Chemistry* vol 9 (Singapore: D. Papousek Ed:
    World Scientific Publishing)
[24] Schmalz T G and Flygare W H. 1978 *Laser and Coherence Spectroscopy* (New York:
    Plenum Press) p. 125ff,
[25] Odashima H, Kajita M, Matsuo Y and Minowa T 1989 *J.Chem. Phys.* **90** 4875
[26] Gierszal S, Galica J and Mis-Kuzminska E 2000 *Physica Scripta* **61** 581
[27] Kajita M, Tachikawa M and Shimizu T 1987 *J. Chem. Phys.* **87** 1620